\documentclass{iopart}
\usepackage{amssymb}
\usepackage{graphicx}
\usepackage{geometry}
\usepackage{lipsum}
\begin{document}

\title[Magnetic Field Effects and Transverse Ratchets in Charge Lattices...]{Magnetic Field Effects and Transverse Ratchets in Charge Lattices Coupled to Asymmetric Substrates}
\author{
C. J. O. Reichhardt and C. Reichhardt
} 
\address{
Theoretical Division and Center for Nonlinear Studies,
Los Alamos National Laboratory, Los Alamos, New Mexico 87545, USA
}
\ead{cjrx@lanl.gov}

%Max 300 words in abstract
\begin{abstract}
We examine a charge lattice coupled
to a one-dimensional asymmetric potential
in the presence of an applied magnetic field, which induces gyrotropic effects in the charge motion. This system could be realized for Wigner crystals in nanostructured samples, dusty plasmas, or other classical charge-ordered states where gyrotropic motion and damping can arise. For zero magnetic field, an applied external ac drive can produce a ratchet effect in which the particles move along the easy flow direction of the substrate asymmetry. The zero field ratchet effect can only occur when the ac drive is aligned with the substrate asymmetry direction; however, when a magnetic field is added, the gyrotropic forces generate a Hall effect that leads to a variety of new behaviors, including a transverse ratchet motion that occurs when the ac drive is perpendicular to the substrate asymmetry direction. We show that this system exhibits commensuration effects as well as reversals in the ratchet effect and the Hall angle of the motion. The magnetic field also produces a nonmonotonic ratchet efficiency when the particles become localized at high fields. 
\end{abstract}

\maketitle

\vskip 2pc

\section{Introduction}

When overdamped particles are coupled to an asymmetric substrate,
a ratchet effect can occur when an ac drive is applied,
leading to a net dc transport in the direction of the
substrate asymmetry.
Ratchet effects have been studied for various
systems in the single particle limit
\cite{Magnasco93,Astumian94,Reimann02},
including colloidal systems \cite{Rousselet94,Marquet02},
biological systems \cite{Lau17}, active matter
\cite{Reichhardt17a}, granular matter \cite{Farkas02},
cold atoms \cite{MenneratRobilliard99,Lundh05},
superconducting vortices \cite{Lee99,Wambaugh99} and
quantum systems \cite{Salger09}.
For a single overdamped particle, the ratchet effect produces
directed motion in the easy direction of
the substrate asymmetry.
When collective interactions become important,
new behaviors emerge,
including commensuration effects
that appear when the spacing between the particles
matches the substrate length scale.
As a result, peaks or dips appear in the ratchet efficiency,
as observed in interacting disk systems \cite{Derenyi95},
active matter systems \cite{McDermott16}
and superconducting vortices
\cite{Villegas03,deSouzaSilva06a,Dinis07a,Lu07,Yu07,Gillijns07,Lin11,VandeVondel11,Shklovskij14,Reichhardt15,Dobrovolskiy20,Lyu21}.
Collective effects can also produce ratchet reversals
in which the net dc flux is in the hard direction of the
substrate asymmetry
\cite{Villegas03,deSouzaSilva06a,Dinis07a,Lu07,McDermott16}.
In many of these systems,
the interactions are short-range; however, collective ratchet effects
can also occur in systems with longer-range interactions where the
particles would naturally form a
triangular lattice in the absence of a substrate,
such as
vortices
in type-II superconductors \cite{Villegas03,Lu07} or
magnetic colloids \cite{Tierno14}.
Ratchet effects can also appear in
charged lattices, which arise in
solid-state systems including Wigner crystals
\cite{Wigner34,Monceau12,Falson22,Shayegan22},
charged colloids with weak screening \cite{Russell15},
optically trapped ions \cite{Schmidt18}, and dusty plasmas \cite{Li23}.

Charged systems form crystalline states when the
Coulomb energy dominates over
the thermal or kinetic energy,
and they can exhibit ratchet effects when coupled to
an asymmetric substrate \cite{Li23,Reichhardt23}.
Little is known, however, about how the ratchet effect is modified
when a magnetic field is applied that causes the moving charges
to undergo cyclotron-like displacements \cite{Brown85,Muller92}.
Previous work on so-called electron pinball states showed
that cyclotron motion can strongly influence the transport of
charged systems coupled to a periodic substrate, resulting in
electron localization
when the cyclotron orbits become
commensurate with the periodicity of the substrate or permitting
delocalized and chaotic motion for incommensurate orbits
\cite{Weiss91,Weiss93}.
When a dc drive is applied in addition to the magnetic field,
the particles move with a finite Hall angle,
and in the presence of
a two-dimensional (2D) periodic substrate,
the motion becomes locked to specific symmetry directions of the substrate,
producing a quantization of the
Hall angle as a function of the magnetic field \cite{Wiersig01,Khoury08}.
Recently, it was shown that when a Wigner crystal in a magnetic field is driven over random disorder, the sliding dynamics occur at a finite
Hall angle that is drive dependent since
the charges are more strongly scattered
by the disorder at low drives \cite{Reichhardt21}.

For particles coupled to one-dimensional (1D) asymmetric substrates, a
ratchet effect only occurs when the ac drive is
applied parallel to the substrate asymmetry direction;
however, once a Hall effect and cyclotron motion are introduced,
it should be possible for a ratchet effect
to appear even for ac drives applied perpendicular to the
substrate asymmetry direction, with the ratcheting
motion consisting of a combination of motion parallel and perpendicular to
the drive.
In magnetic skyrmions, where the Magnus force can have a strong impact
\cite{Nagaosa13,Reichhardt22a} and can
cause the skyrmion motion to obey the same dynamics as
charges in magnetic fields
\cite{Schulz12,Nagaosa13},
there have been several studies showing that new kinds of gyrotropic-induced ratchet effects arise \cite{Reichhardt15,Ma17,Gobel21,Reichhardt22}.
A transverse skyrmion ratchet was also observed
in which the ac driving can be applied at any angle with respect to
the substrate asymmetry direction \cite{Reichhardt15a,Ma17}.
Since the chiral motion itself breaks a symmetry, ratchet effects
in skyrmion systems
can even occur when the substrate is symmetric \cite{Chen20}.
Ratchet effects have been found
in other systems with chiral motion,
even when the substrates are symmetric \cite{Ai16,Liu18,Li23}.
This suggests that similar gyrotropic ratchet effects can
arise in non-skyrmion systems of charged particles in a magnetic field.

Of the numerous classical charged systems that form lattice states
due to Coulomb interactions, the best known is
Wigner crystals of 2D electrons at low densities
\cite{Wigner34,Falson22,Shayegan22}, which
can form in elections on liquid helium \cite{Crandall71,Grimes79}
or in solid-state systems
\cite{Doman79,Bello81,Dykman81,Andrei88,Goldman90,Jiang91,Williams91,Kopelevich07,Monceau12,Brussarski18,Hossain22}.
There have been several studies of 2D Wigner crystals
coupled to 1D or quasi-1D nanostructured channels or arrays
\cite{Shirahama95,Piacente05,Araki12,Rees12,Rees16,Moskovtsev20}, while
more recently, Wigner crystals coupled to
2D periodic substrates have been studied in
moir{\' e} heterostructures
\cite{Regan20,Xu20,Huang21,Li21,Matty22,Mak22,Chen23} and
dichalcogenide monolayers \cite{Smolenski21}.
Under an applied driving force, Wigner crystals can
exhibit sliding states and nonlinear transport
\cite{Andrei88,Goldman90,Jiang91,Williams91,Cha94,Reichhardt01,Kopelevich07,Csathy07,Brussarski18,Hossain22,Reichhardt22}.
There have also been studies examining diode \cite{Zakharov19}
and ratchet effects \cite{Reichhardt23} for Wigner
crystals coupled to asymmetric substrates in the absence of a magnetic field.
Other charged systems that could be coupled to asymmetric substrates include
dusty plasmas \cite{He20,Li23} and ion crystals \cite{Schmidt18}.
Point vortices in fluids with long-range interactions
are undamped but otherwise similar
to charges in a magnetic field,
and under circumstances where damping is present in the fluid, the
point vortices
also exhibit a finite Hall angle under
an applied drift force \cite{Reichhardt20b}.

Previous work on skyrmions interacting
with an asymmetric 1D substrate focused
only on the single skyrmion limit \cite{Reichhardt15},
so collective effects on the motion of gyrotropic particles
over an asymmetric substrate have not yet been addressed.
Additionally, the interactions for skyrmions
generally involve either short-range or intermediate repulsion.
Open questions include how 
a strongly interacting system with gyrotropic
dynamics would behave on an asymmetric substrate,
and what impact collective effects would have on the behavior.

Here, we consider a 2D assembly of charged particles
in the presence of an asymmetric 1D substrate under an ac drive
applied either parallel or perpendicular
to the substrate asymmetry direction.
We specifically study the effect of an applied
magnetic field, which creates velocity components
that are perpendicular to the net force experienced by a charged particle.
When the magnetic field is zero,
a ratchet effect occurs only when the ac drive is
applied parallel to the substrate asymmetry direction,
as found in previous work \cite{Reichhardt23}. 
For finite magnetic fields,
we find that ratchet effects can occur when the ac driving is either
parallel or perpendicular to the substrate asymmetry direction. 
In general, the direction of motion
is at an angle with respect to the direction of the ac drive. 
The efficiency of the
transverse and other magnetic ratchet effects is
generally non-monotonic as a function of field since
the moving charges become
localized at higher fields and execute
small cyclotron orbits that remain confined within a single pinning trough.
We also observe commensuration effects
that arise when the cyclotron orbit radius matches
the substrate periodicity,
as well as a number of ratchet reversals that produce
a reversal in the Hall angle.
In addition to providing examples of new types of
ratchet effects, our results suggest that a transverse ratchet
may be used to detect the presence of Wigner crystals in a magnetic field.

\section{Simulation}
We consider a two-dimensional system with periodic
boundary conditions in the $x$ and $y$ directions
containing $N_e$ particles
with repulsive Coulomb interactions.
The particles also interact with a one-dimensional asymmetric substrate.
The sample is of size $L \times L$ with $L=36$ and the
particle density is $\rho=N_e/L^2$.
The overdamped equation of motion for charge $i$ is given by
\begin{equation}
        \alpha_d {\bf v}_{i} = \sum^{N_e}_{j}\nabla U(r_{ij}) + q{\bf B}\times {\bf v}_{i}
        + {\bf F}_{\rm sub} + {\bf F}_{AC} + {\bf F}^{T} \ .
\end{equation}
Here  $\alpha_{d}$ is the damping constant,
the particle-particle interaction potential is $U(r_{ij}) = q/r_{ij}$,
${\bf r}_i$ and ${\bf r}_j$ are the positions of particles $i$ and $j$,
respectively, $r_{ij}=|{\bf r}_i-{\bf r}_j|$ is the distance
between particles, and  $q$ is the
particle charge which we set to unity.
For computational efficiency, we employ a Lekner method for
calculating the long-range Coulomb interactions,
as used in previous work on charged particles interacting
with a substrate \cite{Lekner91,GronbechJensen97a,Reichhardt01}.
The second term on the right hand side of Eq. (1) describes the Magnus force
produced by a finite magnetic field ${\bf B}=B{\bf \hat{z}}$
applied perpendicular to our 2D sample.

The substrate force ${\bf F}_{\rm sub}=\nabla U(x_i)$ is determined by
the $x$ coordinate of particle $i$, where the
substrate potential $U(x)$
has the form
\begin{equation}
U(x) = -U_{0}[\sin(2\pi x/a) + 0.25\sin(4\pi x/a)] \ .
\end{equation}
This potential produces an asymmetric force profile
with a larger force in the negative $x$-direction.
We measure forces in terms of the parameter $A_p = U_{0}/2\pi$,
and we obtain a maximum force in the hard or negative $x$ direction
of
$F^{\rm hard}_{p} = 1.5A_{p}$
and a maximum force in the easy or positive $x$ direction of
$F^{\rm easy}_{p} = 0.75A_{p}$.
We have previously used this potential to study Wigner crystal ratchet
effects in the absence of a magnetic field \cite{Reichhardt23}.
We concentrate on a system size of $L = 36$
with a substrate lattice spacing of $a=2.117$.    

The ac driving force has the form
${\bf F}_{AC}=  F_{AC}\sin(\omega t){\hat {\bf \alpha}}$,
where $F_{AC}$ is the amplitude and
$\alpha=x$ or $y$ for driving parallel
or perpendicular to the substrate asymmetry direction, respectively.
We use a time step of 
$dt = 0.005$ and
set $\omega = 0.0000754$,
so our typical ac drive cycle spans
$10^5$ simulation time steps. If we decrease $\omega$ to a lower
frequency we find no changes in the behavior;
thus, our results are of relevance to the low frequency limit.
In Section 5
we consider the effects of thermal fluctuations by including
the term ${\bf F}^T$,
which represents Langevin kicks with the properties
$\langle {\bf F}_i^{T}\rangle = 0$ and 
$\langle {\bf F}^T_i(t)\cdot {\bf F}_j^{T}(t^\prime)\rangle = 4\alpha_dk_BT\delta_{ij}\delta(t-t^\prime)$.

\begin{figure}
  \centering
\includegraphics[width=0.8\columnwidth]{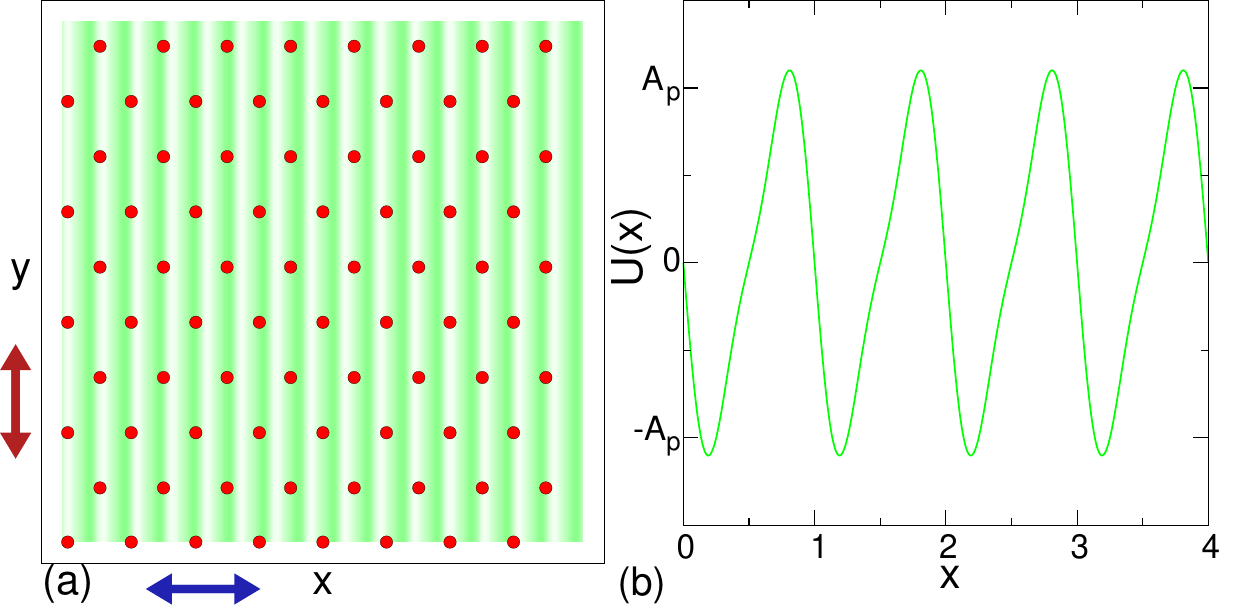}
\caption{
(a) Image of the simulated 2D assembly of charged particles (red circles)
interacting with an asymmetric 1D substrate (green shading). An ac drive
is applied along the $x$ direction (blue arrow), parallel to the substrate
asymmetry direction, or along the $y$ direction (red arrow), perpendicular
to the substrate asymmetry direction.
(b) Detail of the substrate potential $U(x)$ for a lattice constant of
$a=1$.
}
\label{fig:1}
\end{figure}

We measure the average velocity per charge in the $x$ and $y$ directions,
$\langle V_x\rangle=N_e^{-1}\sum_i^{N_e} \langle {\bf v}_i \cdot {\hat {\bf x}}\rangle$ and
$\langle V_y\rangle=N_e^{-1}\sum_i^{N_e} \langle {\bf v}_i \cdot {\hat {\bf y}}\rangle$,
where the time average is taken over 50 ac drive cycles.
We also measure
$|\langle V\rangle|=\sqrt{\langle V_x\rangle^2 + \langle V_y\rangle^2}$.
When $B \neq 0$ and a magnetic field is present,
the charges move at a Hall angle with respect to the drive,
which we measure according to
$\theta_{H} = \arctan(\langle V_{y}\rangle/\langle V_{x}\rangle)$.
The system also has an intrinsic Hall angle,
given by $\theta^{\rm int}_H = \arctan(qB/\alpha_{d})$, which increases
with increasing $B$. Interactions with the substrate can cause
$\theta_{H}$ to be different from $\theta^{\rm int}_H$,
as shown in previous work \cite{Reichhardt21}.
Wiersig {\it et al.}  \cite{Wiersig01} considered charges
moving over a 2D periodic array of obstacles under applied
magnetic fields that would produce intrinsic Hall angles as large as
$\theta^{\rm int}_H=60^\circ$.
In Fig.~\ref{fig:1}(a), we show an image of the system with arrows
indicating the ac driving direction that is either parallel to the
substrate asymmetry direction, giving a longitudinal ratchet effect,
or perpendicular to the substrate asymmetry direction, giving
a transverse ratchet effect.
Figure~\ref{fig:1}(b) shows a detail of the substrate potential $U(x)$
for a sample where the substrate lattice period is $a=1$.

\section{Results}

\begin{figure}
  \centering
\includegraphics[width=0.6\columnwidth]{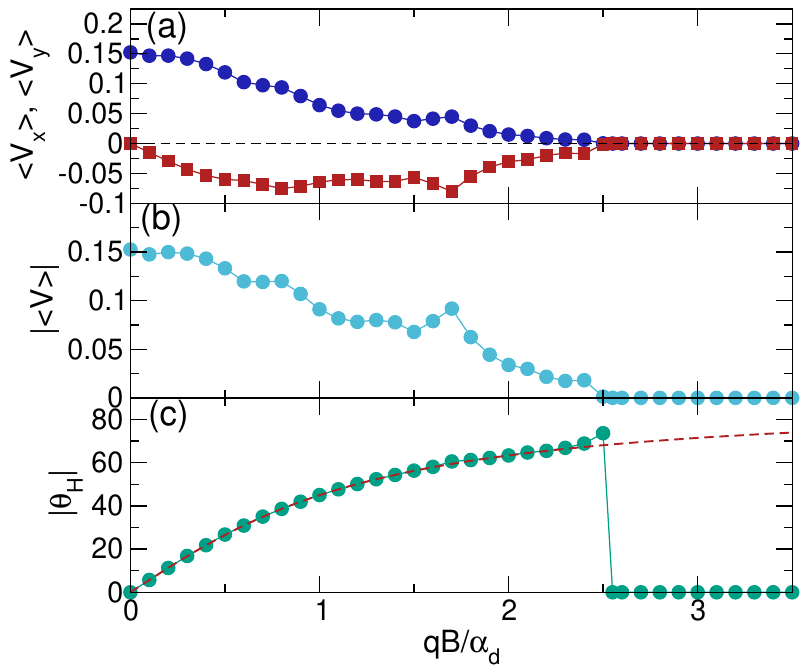}
\caption{ 
(a) $\langle V_{x}\rangle$ (blue) and $\langle V_y\rangle$ (red) vs
$qB/\alpha_d$ for a system with $\rho = 0.208$,
$A_p = 0.75$, and $F_{AC} = 1.0$,
where the ac drive is applied along $x$, parallel
to the substrate asymmetry direction.
(b) The corresponding $|\langle V\rangle|$ vs $qB/\alpha_d$.
(c) The measured Hall angle $|\theta^H|$ vs $qB/\alpha_d$.
The dashed line is a fit to
$\theta^{H} = \arctan(qB/\alpha_d)$.
}
\label{fig:2}
\end{figure}

\begin{figure}
  \centering
\includegraphics[width=0.6\columnwidth]{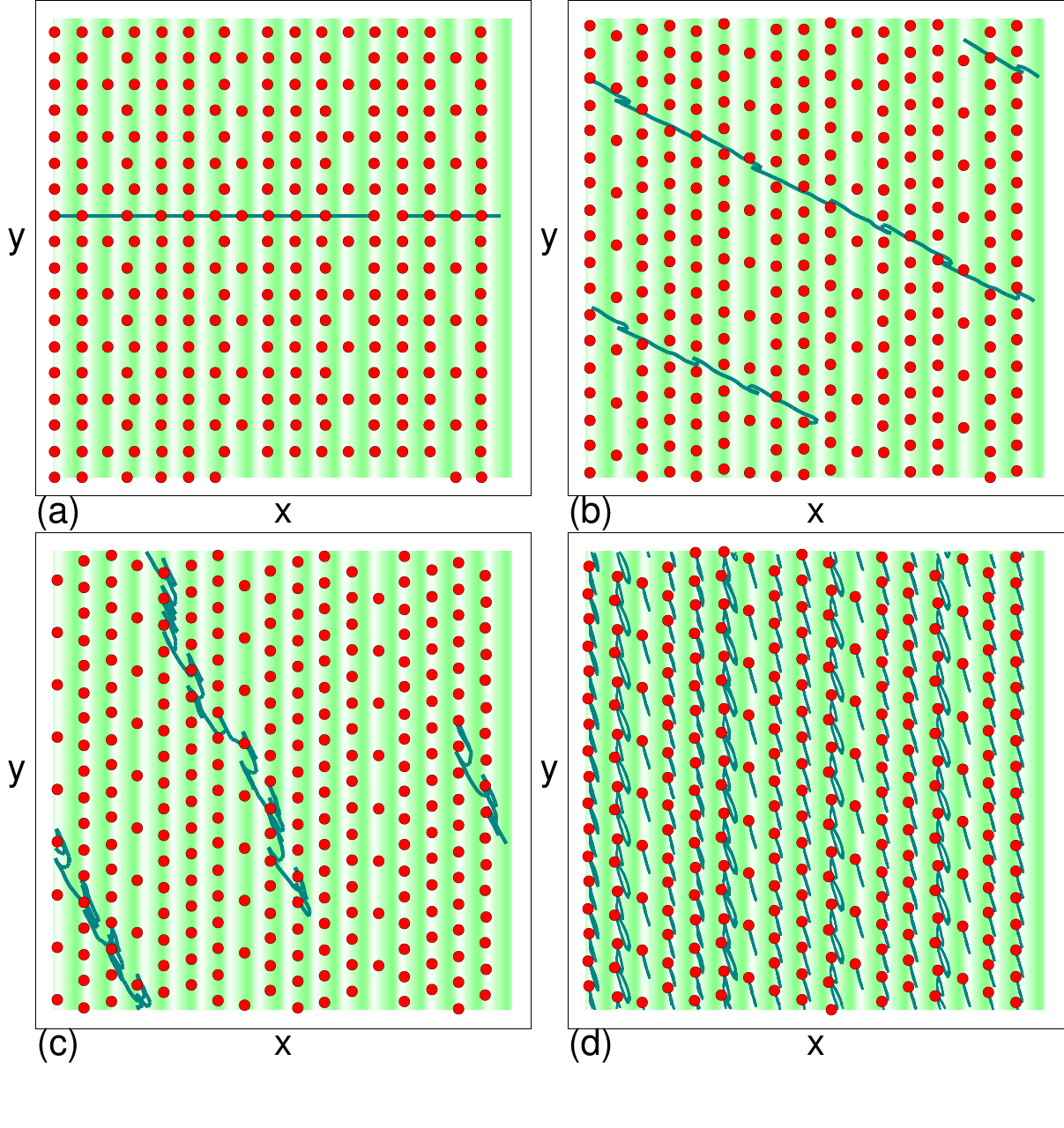}
\caption{ 
(a,b,c) Particle locations (red circles), substrate (green shading),
and the trajectory of a single
representative particle (line) for the system in
Fig.~\ref{fig:2} with $\rho=0.208$, $A_p=0.75$, and $F_{AC}=1.0$ for
driving along $x$, parallel to the substrate asymmetry direction.
(a) $qB/\alpha_d = 0.0$.
(b) $qB/\alpha_d = 0.5$.
(c) $qB/\alpha_d = 2.0$.
(d) Trajectories of all particles for
$qB/\alpha_d=3.5$,
where the motion is localized and there is no ratchet effect.
}	
\label{fig:3}
\end{figure}

In Fig.~\ref{fig:2}(a) we plot $\langle V_{x}\rangle$ and
$\langle V_y\rangle$ versus 
$qB/\alpha_{d}$ for a system with
$\rho = 0.208$, $A_p = 0.75$, and $F_{AC} = 1.0$.
Here, the ac drive is applied along $x$, parallel to the substrate
asymmetry direction.
For this system, the maximum substrate forces are
$F^{\rm hard}_{p} = 1.125$ and $F^{\rm easy}_{p}  = 0.5625$.
When $B = 0.0$, we find a pronounced 
longitudinal ratchet effect with $\langle V_x\rangle \neq 0$,
as studied in previous work \cite{Reichhardt23}.
Figure~\ref{fig:2}(b) shows
the average velocity $|\langle V\rangle|$ versus $qB/\alpha_d$
for the same sample.
In Fig.~\ref{fig:3}(a), we illustrate the particle locations and
substrate along with the trajectory of a representative particle
during 10 ac drive cycles. The same type of trajectory is followed
by all of the other particles, each of which
moves only along the $x$-direction over time with a net motion
occurring in the $+x$ direction. 
In Fig.~\ref{fig:2}(c), we plot
$|\theta_H|$, the absolute value of the
measured Hall angle, versus $qB/\alpha_d$.
As $B$ increases,
$\langle V_{y}\rangle$ becomes nonzero when the particles
begin to move along the negative $y$-direction,
and a finite Hall angle emerges.
With a further increase of $B$,
$|\langle V_y\rangle|$ reaches its maximum value near
$qB/\alpha_d = 0.8$ and then diminishes again.
The overall velocity $|\langle V\rangle|$
generally decreases with increasing $B$.
For $qB/\alpha_d = 0.5$, the particles move close to an angle of
$\theta_H=30^\circ$, as illustrated in Fig.~\ref{fig:3}(b),
while at $qB/\alpha_d = 2.0$,
Fig.~\ref{fig:3}(c) shows that
the particles move along $62^\circ$ since the Hall angle increases with
increasing $B$.
The dashed line in
Fig.~\ref{fig:2}(c) indicates
what the Hall angle would be if
$A_p = 0.0$, as determined by the expression
$\theta_H = \arctan(qB/\alpha_d)$.
Near $qB/\alpha_d = 1.7$, we find a local peak in the velocities,
corresponding to the field at which the
orbit size partially matches the periodicity of the substrate.
For $qB/\alpha_d \geq 2.5$,
the drift velocities drop to zero
because the particle motion
becomes localized, as illustrated in Fig.~\ref{fig:3}(d) where we
highlight the trajectories of all
of the particles.
For larger $B$, the particle orbits become more compact and 
remain confined in a single pinning trough.

\begin{figure}
  \centering
\includegraphics[width=0.6\columnwidth]{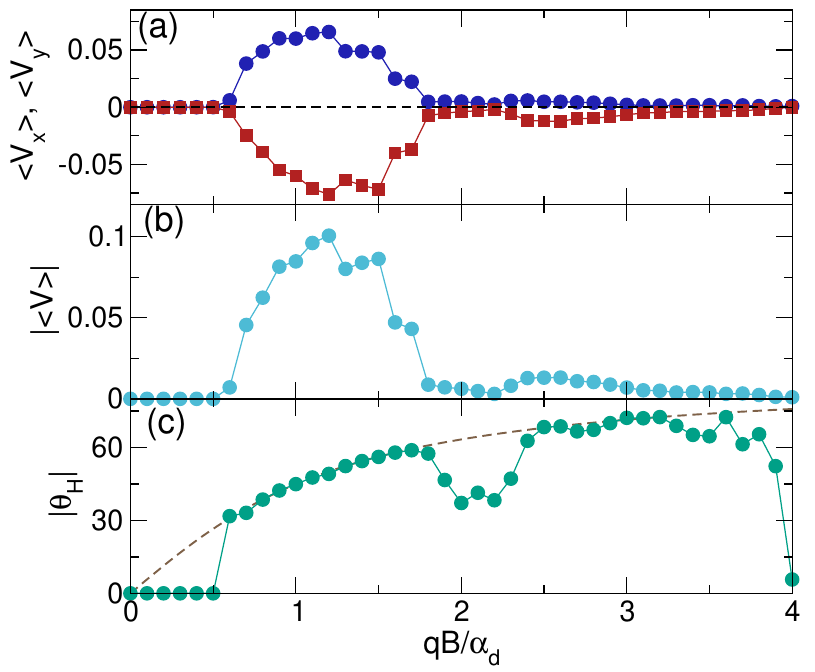}
\caption{
$\langle V_{x}\rangle$ (blue) and $\langle V_y\rangle$ (red) vs
$qB/\alpha_d$ for the system from Fig.~\ref{fig:2}
with $\rho=0.208$, $A_p=0.75$, and $F_{AC}=1.0$
but with the ac drive applied along the $y$-direction,
perpendicular to the substrate asymmetry direction.
(b) The corresponding $|\langle V\rangle|$ vs $qB/\alpha_d$.
(c) $|\theta_H|$ vs $qB/\alpha_d$. The dashed line is a fit to
$\theta_{H} = \arctan(qB/\alpha_d)$.}
      \label{fig:4}
\end{figure}

\begin{figure}
  \centering
\includegraphics[width=0.6\columnwidth]{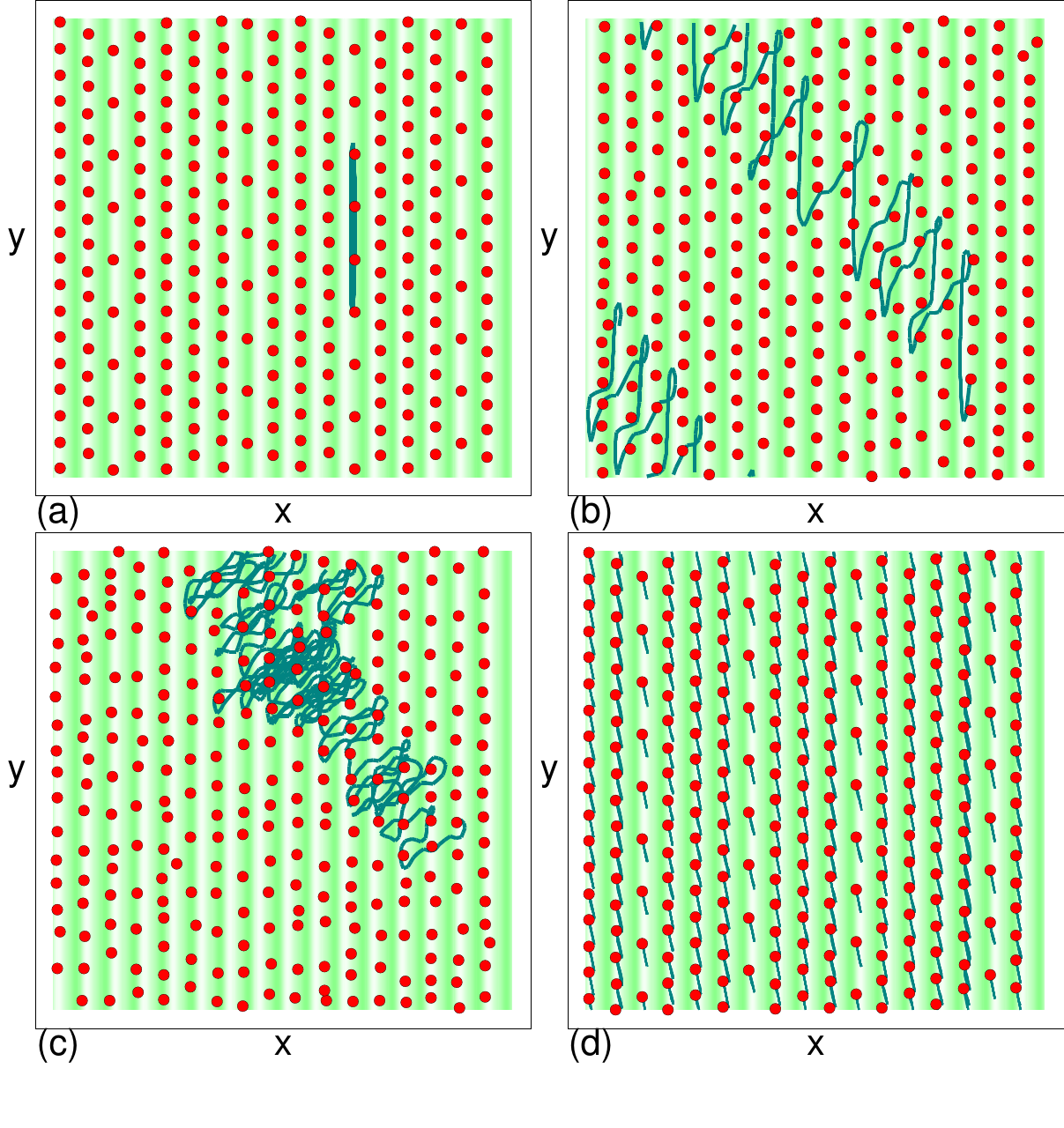}
\caption{
(a,b,c) Particle locations (red circles), substrate (green shading), and
the trajectory of a single representative particle (line) for the
system in Fig.~\ref{fig:2} with $\rho=0.208$, $A_p=0.75$,
and $F_{AC}=1.0$ for driving along $y$, perpendicular to the substrate
symmetry direction.
(a) $qB/\alpha_d = 0.5$ where the motion consists of oscillations
along $y$.
(b)
$qB/\alpha_d = 1.2$, showing a 2D periodic translating orbit.
(c)
$qB/\alpha_d = 2.5$, where the motion
is more disordered or chaotic.
(d) The trajectories for all the particles
at $qB/\alpha_d = 4.0$, where the motion is localized.
}
\label{fig:5}
\end{figure}

In Fig.~\ref{fig:4} we plot $\langle V_{x}\rangle$ and
$\langle V_y\rangle$ versus
$qB/\alpha_d$ for the same system as in
Fig.~\ref{fig:2} but with the ac drive applied perpendicular to the
substrate asymmetry direction.
For $qB/\alpha_d < 0.6$, there is no ratchet effect
and both $\langle V_{x}\rangle$ and $\langle V_{y}\rangle$ are zero.
Figure~\ref{fig:5}(a) shows the trajectory of a single particle
at $qB/\alpha_d = 0.5$, where the particles only move along the $y$ direction
and remain localized.
There is a large transverse ratchet effect
for $0.6 \leq qB/\alpha_d < 1.9$,
a reduced ratchet effect appears for $1.9 \leq qB/\alpha_d < 3.9$,
and for $qB/\alpha_d \geq 3.9$, the motion is localized again.
In Fig.~\ref{fig:5}(b)
at $qB/\alpha_d = 1.2$, the orbit of a single particle consists of
a combination of sliding motion along $y$ and jumps along $x$,
giving a net motion that is in the positive $x$ and negative $y$
direction. For $1.9 \leq qB/\alpha_d < 3.9$, the motion is much more chaotic
and produces a gradual drift, as illustrated
in Fig.~\ref{fig:5}(c) at $qB/\alpha_d = 2.5$.
Figure~\ref{fig:5}(d) shows
that at $qB/\alpha_d = 4.0$, the motion has become
localized within the pinning troughs.
In Fig.~\ref{fig:4}(c) we plot $|\theta^H|$ versus $qB/\alpha_d$
along with a dashed line
that indicates the behavior expected in a substrate-free system.
A dip appears when $1.9 < qB/\alpha_d < 2.5$,
where the motion is primarily chaotic with a reduced velocity,
as is also visible in the plot of
$|\langle V\rangle|$ versus $qB/\alpha_d$
in Fig.~\ref{fig:4}(b).
Our results demonstrate
that a magnetic field can induce a ratchet effect
even when the ac drive is applied perpendicular to the
substrate asymmetry direction, and that this
ratchet effect is non-monotonic as a function of the field.
The maximum net velocity for the transverse ratchet is close to
that of the longitudinal ratchet found for ac driving applied
parallel to the substrate asymmetry direction.
We find that there is a critical field
that must be applied in order for the traverse ratchet to occur,
and that the value of this field depends on the strength of the substrate.

\begin{figure}
  \centering
\includegraphics[width=0.6\columnwidth]{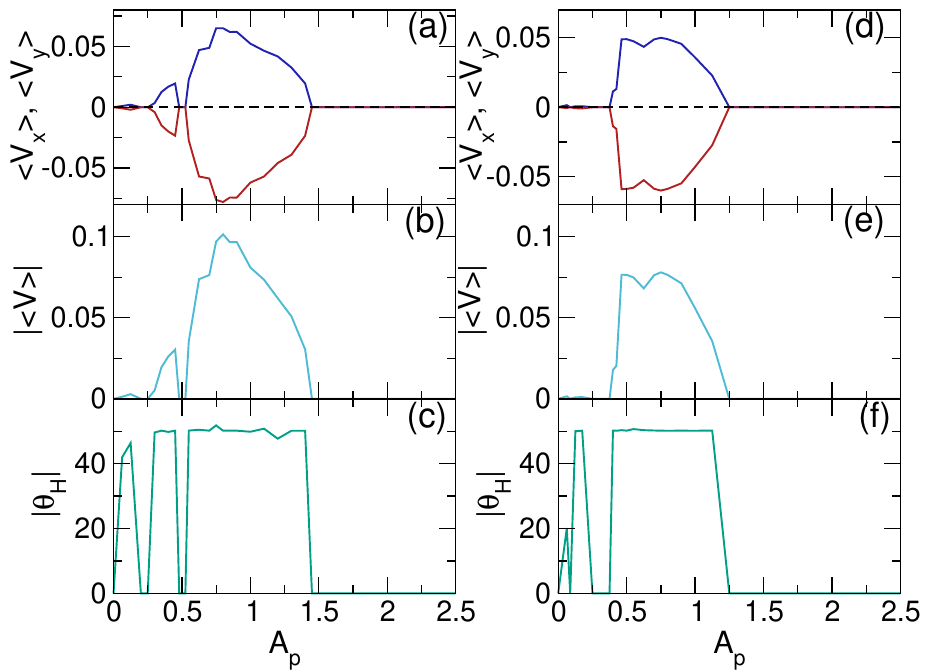}
\caption{
(a) $\langle V_{x}\rangle$ (blue) and $\langle V_{y}\rangle$ (red) vs
$A_{p}$ for the system in Fig.~\ref{fig:2}
with $\rho=0.208$ and $F_{AC}=1.0$ at $qB/\alpha_d = 1.2$
for ac driving applied along $y$ direction, perpendicular to the
substrate asymmetry direction.
(b) The corresponding $|\langle V\rangle|$ vs $A_{p}$.
(c) The corresponding $|\theta_{H}|$ vs $A_p$.
(d) $\langle V_x\rangle$ (blue) and $\langle V_{y}\rangle$ (red) vs
$A_{p}$ for the same system with ac driving
applied along $x$, parallel to the substrate asymmetry
direction.
(e) The corresponding $|\langle V\rangle|$ vs $A_{p}$.
(f) The corresponding $|\theta_{H}|$ vs $A_{p}$.
}
\label{fig:6}
\end{figure}

In Fig.~\ref{fig:6}(a), we plot $\langle V_{x}\rangle$ and
$\langle V_{y}\rangle$ versus $A_{p}$ for the same system from
Fig.~\ref{fig:2}
with $F_{AC} = 1.0$ and  $qB/\alpha_d = 1.2$
under driving perpendicular to the
substrate asymmetry direction.
Figure~\ref{fig:6}(b) and (c) show the corresponding
$|\langle V\rangle|$ and $|\theta_H|$ versus $A_p$.
There is no ratchet effect at low $A_{p}$ since the system forms a
lattice that effectively floats above the substrate.
We find several regions where the ratchet effect goes to zero
because the particle orbits become localized.
For $A_p > 1.5$, the system is pinned.
In Fig.~\ref{fig:6}(c), the Hall angle in the ratcheting
regimes is close to the value expected for a pin-free system.
Figure~\ref{fig:6}(d,e,f) shows $\langle V_x\rangle$,
$\langle V_y\rangle$, $|\langle V\rangle|$, and $|\theta_H|$ versus
$A_p$ for the same system but for ac driving parallel to the
substrate asymmetry direction. Here the ratchet effect
is lost for both low and high $A_{p}$ values.

\begin{figure}
  \centering
\includegraphics[width=0.6\columnwidth]{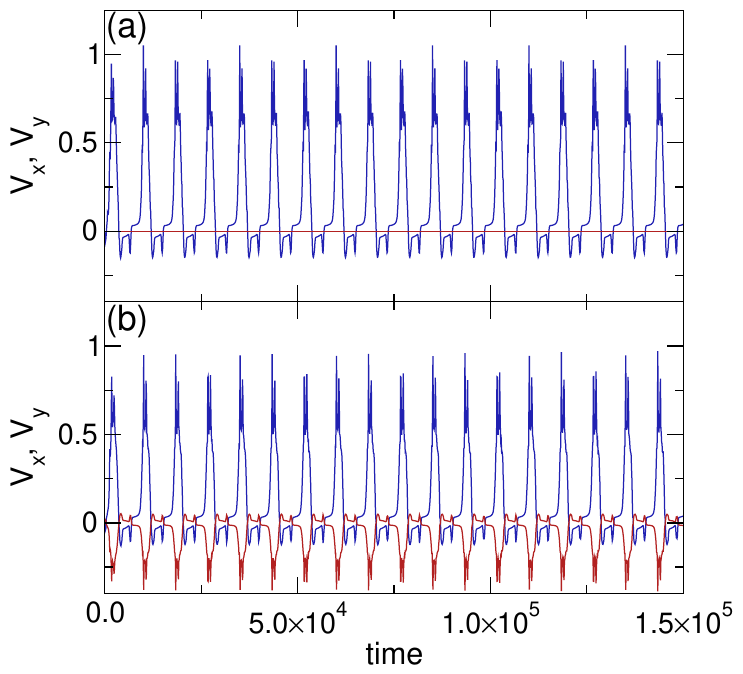}
\caption{
Time series of instantaneous velocities $V_{x}$ (blue) and $V_{y}$ (red)
for the system in Fig.~\ref{fig:2}
with $\rho=0.208$, $A_p=0.75$, and $F_{AC}=1.0$
for driving along $x$, parallel to the substrate asymmetry
direction.
(a) $qB/\alpha_d = 0.0$, where there is ratcheting motion
in the positive $x$ direction but no motion in the $y$ direction. 
(b) $qB/\alpha_d = 0.4$, where there is a strong ratcheting
motion in the $x$-direction
and a smaller ratcheting motion in the negative $y$-direction.
}
\label{fig:7}
\end{figure}

\begin{figure}
\centering
\includegraphics[width=0.6\columnwidth]{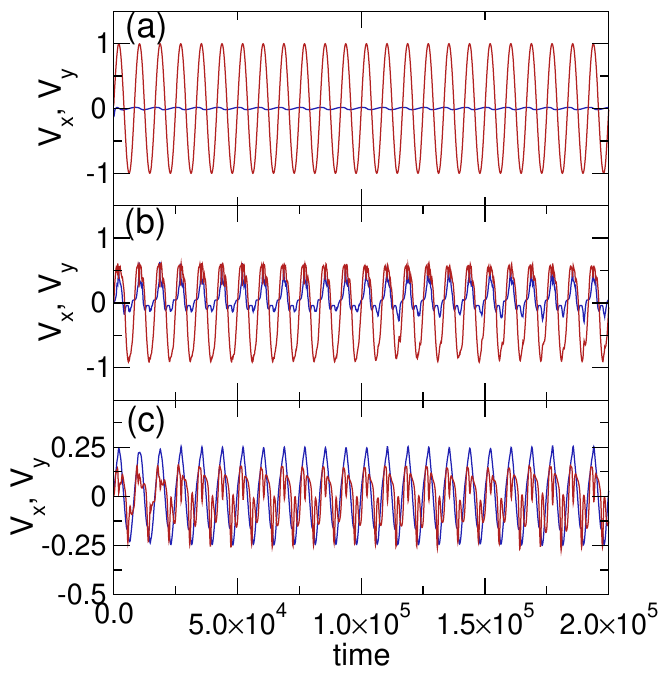}
\caption{
Time series of instantaneous velocities  
$V_{x}$ (blue) and $V_{y}$ (red)
for the system in Fig.~\ref{fig:2}
with $\rho=0.208$, $A_p=0.75$, and $F_{AC}=1.0$
for driving along $y$,
perpendicular to the substrate asymmetry direction.
(a) At $qB/\alpha_d = 0.4$, there is no ratcheting motion in either direction.
(b) At $qB/\alpha_d = 1.2$, there is ratcheting motion in the
positive $x$-direction and the negative $y$-direction.
(c) At $qB/\alpha_d = 4.0$, there is no ratcheting motion in either direction.
}
\label{fig:8}
\end{figure}

In Fig.~\ref{fig:7}(b), we plot time series of the instantaneous
values of $V_{x}$ and $V_{y}$ (red) for the system in
Fig.~\ref{fig:2} at $qB/\alpha_d = 0.4$ for
ac driving applied parallel to the substrate asymmetry direction,
where we find velocity spikes in the positive $x$ direction and 
smaller spikes in the negative $y$ direction.
The additional features in the time series correspond to particle
configurations that form during
specific portions of the ac drive cycle, since
the overall structure of the particles
can change as the ac drive direction varies throughout the cycle.
At $qB/\alpha_d = 0.0$, Fig.~\ref{fig:7}(a) shows that
spikes are still present in $V_{x}$ but are absent for $V_{y}$.
In Fig.~\ref{fig:8}, we plot the instantaneous
$V_{x}$ and $V_{y}$ versus time for the same system with ac driving
applied perpendicular to the substrate asymmetry direction.
At $qB/\alpha_d = 0.4$ in Fig.~\ref{fig:8}(a),
there are strong oscillations in $y$ but
the oscillations are symmetric so
$\langle V_{y}\rangle = 0$.
There are much smaller oscillations in $x$
but these oscillations are also symmetric and
$\langle V_{x}\rangle = 0$.
In Fig.~\ref{fig:8}(b) at $qB/\alpha_d = 1.2$, there are strong spikes in $V_x$
in the positive $x$ direction corresponding to net motion of the particles
in this direction. Both positive and negative velocity spikes appear
in $V_y$, but the spikes are stronger in the negative $y$ direction, leading
to ratcheting motion along $-y$.
Figure~\ref{fig:8}(c) shows that at $qB/\alpha_d = 4.0$,
there are velocity oscillations in both directions
but $\langle V_{x}\rangle$ and $\langle V_{y}\rangle$ are both zero.

\section{Ratchet Effects for Varied ac drives}

\begin{figure}
\centering
\includegraphics[width=0.6\columnwidth]{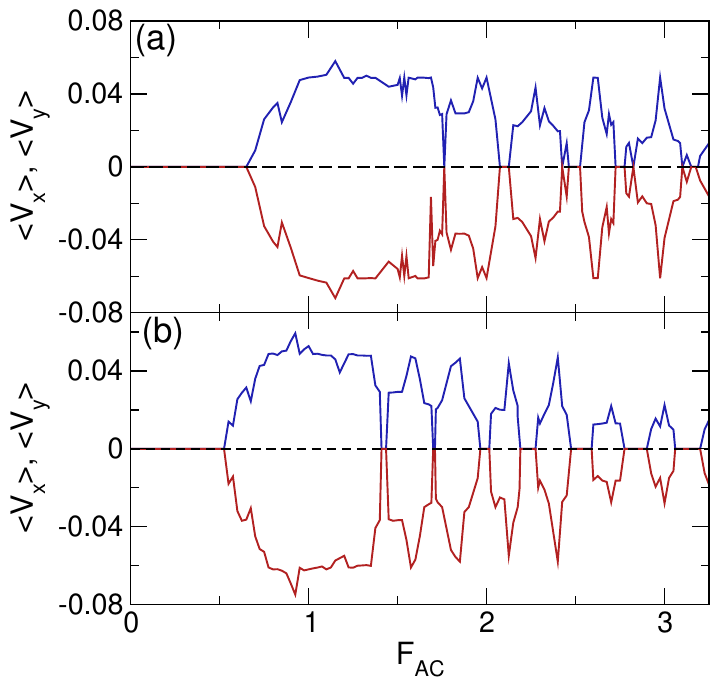}
\caption{
(a) $\langle V_{x}\rangle$ (blue) and $\langle V_{y}\rangle$ (red)
vs $F_{AC}$ for the system from Fig.~\ref{fig:6} with
$\rho=0.208$ and $F_{AC}=1.0$
at $A_{p} = 0.75$ and $qB/\alpha_d = 1.2$ for driving along $x$,
parallel to the substrate asymmetry direction.
(b) The same for driving along $y$, perpendicular to the
substrate asymmetry direction.
In both cases, there are regions where
the ratchet efficiency drops to zero.
}
\label{fig:9}
\end{figure}

In Fig.~\ref{fig:9}(a) we plot $\langle V_x\rangle$ and
$\langle V_y\rangle$ versus $F_{AC}$ for the system in
Fig.~\ref{fig:6} at $A_{p} = 0.75$ and $qB/\alpha_d = 1.2$
for driving parallel to the substrate asymmetry direction.
Here, the velocities are zero for $F_{AC} < 0.6$
because the ac drive is not large enough to permit the particles
to jump out of the the wells.
For $0.6 \leq F_{AC} < 1.8$, there is a strong ratchet effect where the
particles move in the positive $x$ and negative $y$
directions, giving
$\theta_H \approx 55^\circ$.
We find a series of regions where the ratchet effect
drops to zero, corresponding to ac drive amplitudes
where the particle orbits become
trapped inside the pinning troughs.
In other regions, the ratchet effect is reduced but there is still
finite ratcheting motion
corresponding to orbits
in which the particles move one lattice constant every other ac cycle or
every $n$th ac cycle.
For ac driving applied perpendicular to the substrate asymmetry direction,
Fig.~\ref{fig:9}(b) shows a similar pattern of regions of reduced or
zero ratcheting;
however, the $F_{AC}$ values at which zero velocities appear
are shifted relative to the parallel driving case since
the particle orbits are more tilted under perpendicular driving.
Behavior similar to that shown in Fig.~\ref{fig:9},
with oscillations in the ratchet effectiveness as a function of ac amplitude,
was observed previously in purely overdamped systems
such as superconducting vortices on asymmetric 1D
substrates \cite{Lee99}.
In general, we find that the oscillations in ratchet efficiency
are the most pronounced for low particle densities.
At higher densities, oscillations appear for fillings at which the system is relatively ordered, while for incommensurate fillings,
regions where the ratchet effect drops to zero are
replaced with regions of finite but reduced ratchet effect. 

\begin{figure}
\centering
\includegraphics[width=0.6\columnwidth]{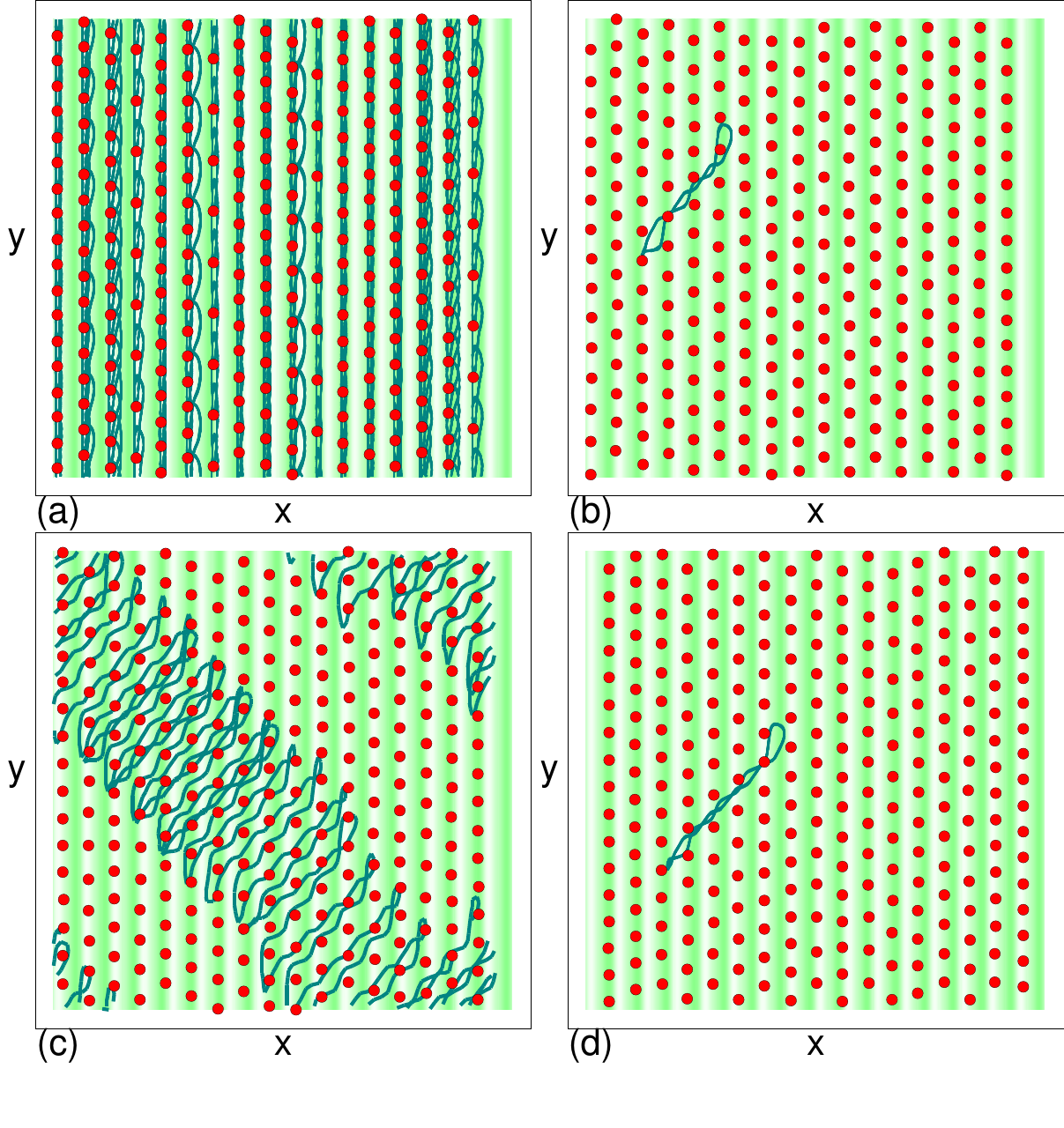}
\caption{
(a) Particle locations (red circles), substrate (green shading), and the
trajectories of all the particles (lines) for the system in Fig.~\ref{fig:9}(b)
with $\rho=0.208$, $A_p=0.75$, and
$qB/\alpha_d=1.2$ for driving along $y$,
perpendicular to the substrate asymmetry direction, at
$F_{AC} = 0.5$ where there is no ratchet effect.
(b,c,d) The same but with the trajectory of only a single representative
particle shown.
(b)
$F_{AC} = 1.41$, where there is no ratchet effect
and the particle orbit is localized.
(c) $F_{AC} = 1.6$,
where there is a finite ratchet effect.
(d) $F_{AC} = 1.7$, where
the ratchet effect is absent and the particle motion is localized.
}
\label{fig:10}
\end{figure}    

In Fig.~\ref{fig:10}(a), 
we illustrate the particle locations and trajectories
for the system in Fig.~\ref{fig:9}(b) for 
driving along $y$,
perpendicular to the substrate asymmetry direction,
at $F_{AC} = 0.5$ where the ratchet effect is zero. 
In this case, the particles oscillate in closed orbits,
and there is no hopping from one well to the next.
Figure~\ref{fig:10}(b) shows the trajectory of a single
representative particle
at $F_{AC} = 1.41$
where $\langle V_x\rangle = \langle V_y\rangle = 0$.
Here each particle moves in a closed orbit
spanning three pinning potential minima.
At $F_{AC}=1.6$, plotted in Fig.~\ref{fig:10}(c),
there is a finite ratchet effect.
Here, each particle follows a ratcheting trajectory that
covers close to 3.5 pinning troughs per orbit.
In Fig.~\ref{fig:10}(d) at $F_{AC} = 1.7$,
the particles form closed orbits, and
Fig.~\ref{fig:9}(b) indicates that there is zero ratchet effect.
In this case, each particle circulates among
four pinning minima.
The ratchet effect is generally lost in orbits where the particles are confined between $n$ pinning minima, while there is a finite ratchet effect
for incommensurate orbits.
For driving along $x$,
parallel to the substrate asymmetry direction,
a similar trapping effect occurs for regions where
the ratchet effect is zero;
however, the particle orbits are more one-dimensional in character
compared to the perpendicular driving case.

\subsection{Ratchet Reversals for Varied Filling and Substrate Strength}

\begin{figure}
\centering
\includegraphics[width=0.6\columnwidth]{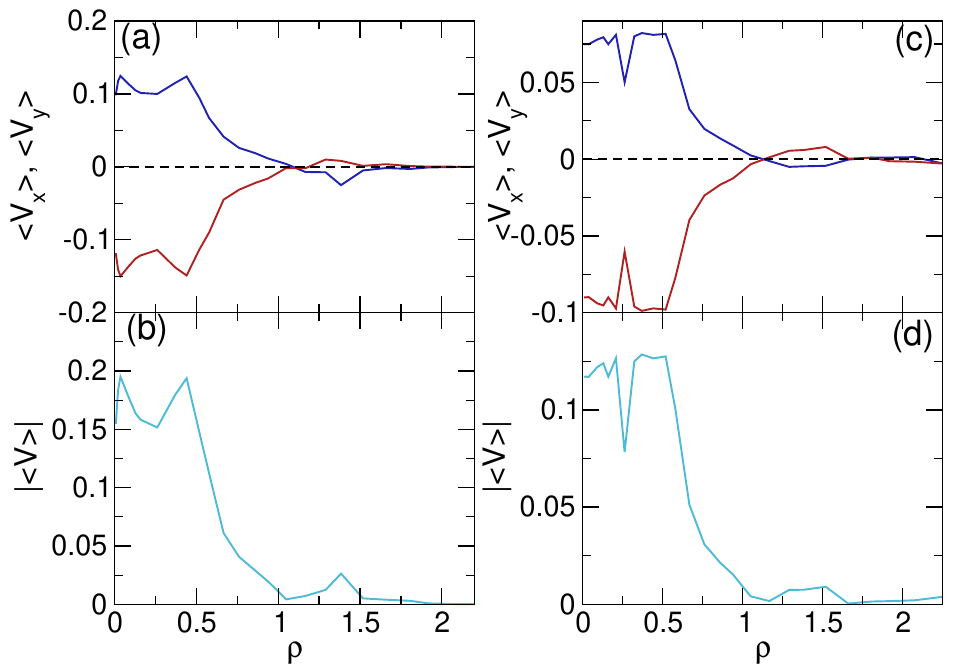}
\caption{
(a) $\langle V_{x}\rangle$ (blue) and $\langle V_{y}\rangle$ (red) vs
particle density $\rho$ for a system with 
$qB/\alpha_d=1.2$,
$A_p = 1.25$, and $F_{AC} = 1.5$ for driving along $x$,
parallel to the substrate asymmetry direction.
(b)
The corresponding
$|\langle V\rangle|$ vs $\rho$.
(c) $\langle V_{x}\rangle$ (blue) and
$\langle V_{y}\rangle$ (red) vs $\rho$ for the same system
but for driving along $y$,
perpendicular to the substrate asymmetry direction.
(d) The corresponding $|\langle V\rangle|$ vs $\rho$.
There
is a reversal in the ratchet effect
as a function of increasing $\rho$ for both ac driving directions.
}
\label{fig:11}
\end{figure}
                
We next consider the effect of 
varying the charge density $\rho$.
In Fig.~\ref{fig:11}(a) we plot $\langle V_{x}\rangle$ and
$\langle V_{y}\rangle$ versus $\rho$ for a system with $A_p = 1.25$,
$qB/\alpha_d = 1.2$, and $F_{AC} = 1.5$
for driving perpendicular to the substrate asymmetry direction.
For $\rho < 1.0$, we find a ratchet effect in which
the particles move in the positive $x$ and negative $y$ directions.
This ratchet motion passes through a local maximum near $\rho = 0.5$.
There is a reversal in the ratchet for $\rho > 1.0$,
where the particles move in the positive $y$ and negative $x$ directions.
Figure~\ref{fig:11}(b) shows the absolute value of the velocity
$|\langle V\rangle|$ versus $\rho$, which
peaks near $\rho = 0.5$
and has a drop near $\rho  = 1.0$ at the point where the ratchet
reversal occurs.
In the reversed ratchet regime,
the absolute value of the velocity is much lower
since the motion is produced by a small number of solitons
in the lattice that can hop over the substrate barriers in the
hard direction.
During the $+x$ portion of the ac drive cycle,
the particles form a more spread out configuration
and the solitons are less well defined.
In previous work performed at $qB/\alpha_d = 0.0$,
a reversal of the ratchet effect was also
observed as a function of increasing filling \cite{Reichhardt23}.
In Fig.~\ref{fig:11}(c), we plot $\langle V_{x}\rangle$
and $\langle V_{y}\rangle$ versus $\rho$
for the same system from Fig.~\ref{fig:11}(a)
but with the ac driving applied perpendicular to the substrate asymmetry
direction.
For $\rho < 1.0$, the motion is
in the positive $x$ and negative $y$ directions.
In the corresponding plot of $|\langle V\rangle|$ versus $\rho$ in
Fig.~\ref{fig:11}(d),
there is a dip near $\rho = 0.25$ due to the formation of a commensurate
lattice, which is better pinned by the substrate.
For $\rho > 1.0$, the overall velocity is strongly reduced,
and there is a ratchet reversal at $\rho\approx 1.2$ where the motion is in
the positive $y$ and negative $x$ directions.

\begin{figure}
\centering
\includegraphics[width=0.6\columnwidth]{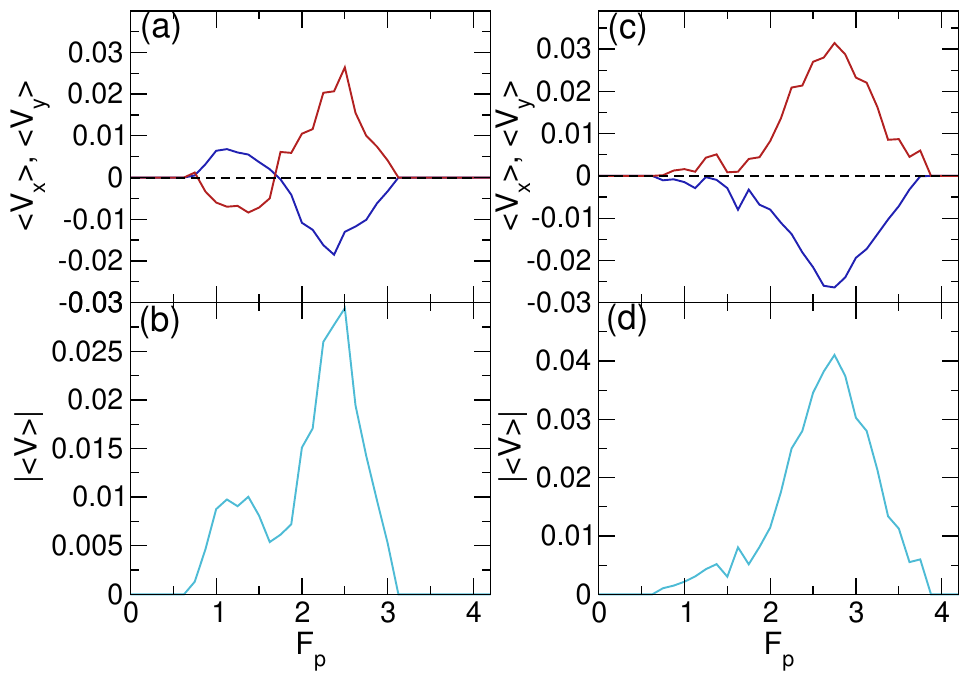}
\caption{ 
(a) $\langle V_{x}\rangle$ (blue) and $\langle V_{y}\rangle$ (red)
vs $F_{p}$ for a sample with $F_{AC} = 1.0$
and $qB/\alpha_d = 1.2$ for driving along $y$, perpendicular to the
substrate asymmetry direction,
at $\rho = 0.938$ where a ratchet reversal occurs.
(b) The corresponding $|\langle V\rangle|$ vs $F_p$.
(c) $\langle V_x\rangle$ (blue) and $\langle V_y\rangle$ (red)
for the same system at $\rho = 1.27$ where there is only a reversed
ratchet effect.
(d) The corresponding $|\langle V\rangle|$ vs $F_p$.
}
\label{fig:12}
\end{figure}

The magnitude and direction of the ratchet motion
depends on both the filling fraction and the substrate strength.
For lower fillings of $\rho < 0.77$,
the ratchet motion is generally in the positive
$x$ direction; however, for higher fillings,
the ratchet motion can be in either the positive or negative
$x$ direction.
In Fig.~\ref{fig:12}(a) we plot $\langle V_{x}\rangle$ and
$\langle V_{y}\rangle$ versus $F_{p}$
for a system with $\rho = 0.938$, $F_{AC} = 1.0$,
and $qB/\alpha_d = 1.2$ under driving along $y$, perpendicular
to the substrate asymmetry, while
Fig.~\ref{fig:12}(b) shows the corresponding
$|\langle V\rangle|$ versus $F_{p}$.
For $F_{p} < 0.6$ there is no ratchet effect.
A positive ratchet effect with motion in the
positive $x$ and negative $y$ directions appears for
$0.6 \leq F_{p} < 1.3$, and is
followed by a ratchet reversal for $1.3 \leq F_{p} < 3.0$,
where the motion is in the positive $y$ and negative $x$
direction.
When the ratchet reverses, the particle motion reverses
direction.
All of the particles are pinned for $F_{p} \geq 3.0$.
Both the forward and reverse ratchet effects pass through peak
efficiencies, leading to the appearance of
a double peak feature in $|\langle V\rangle|$.
Figure~\ref{fig:12}(c,d) shows $\langle V_x\rangle$, $\langle V_y\rangle$,
and $|\langle V\rangle|$ versus $F_p$
for the same system at a higher density of
$\rho=1.27$,
where only a reversed ratchet effect occurs.

\begin{figure}
\centering
\includegraphics[width=0.6\columnwidth]{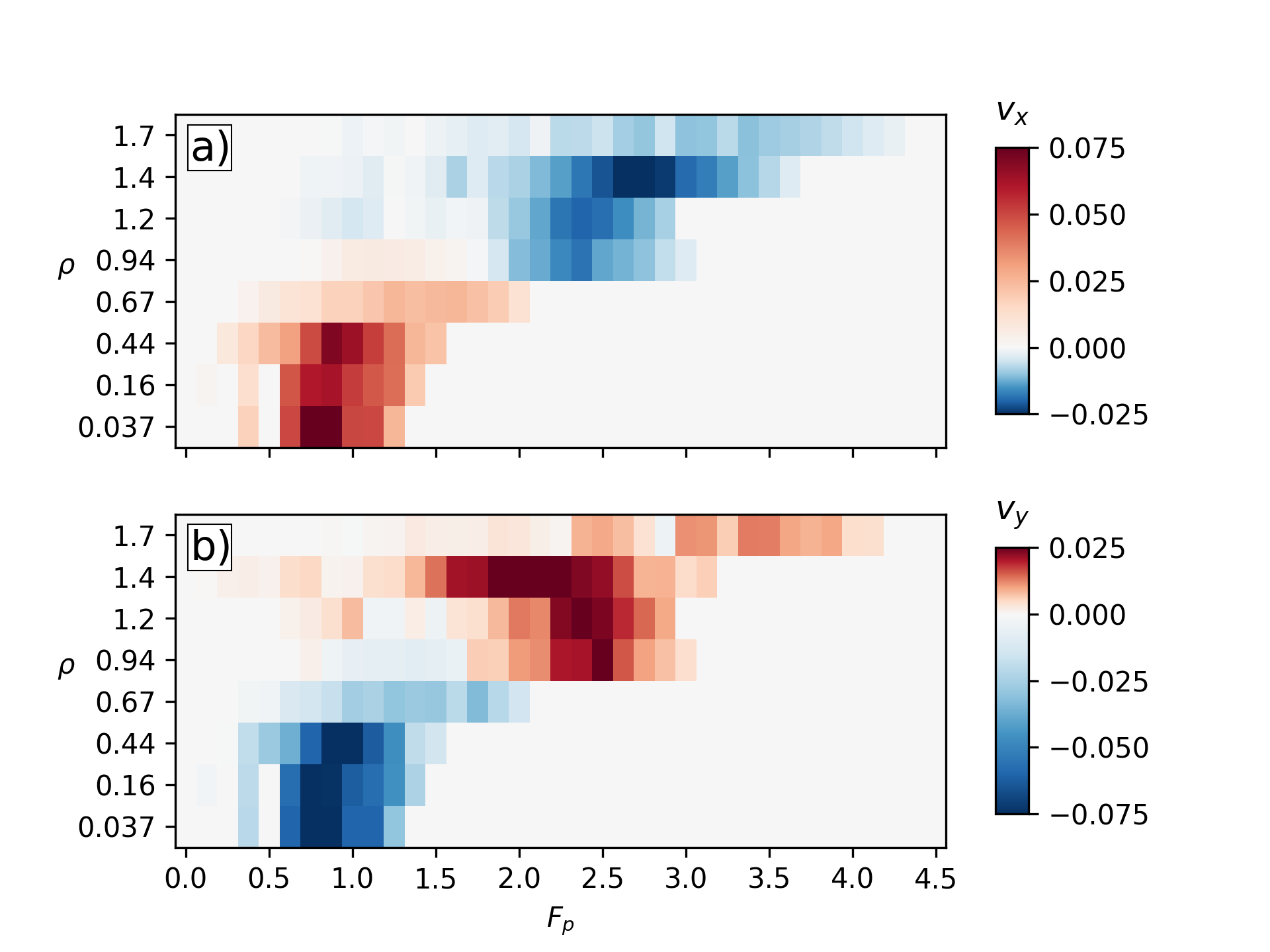}
\caption{
Heat map of the ratchet velocity as a function of  
$\rho$ vs $F_{p}$ for the system in Fig.~\ref{fig:12} with
$F_{AC}=1.0$ and $qB/\alpha_d=1.2$ for driving along $y$,
perpendicular to the substrate asymmetry direction.
(a) $\langle V_{x}\rangle$. (b) $\langle V_{y}\rangle$.
The change in color from red tones to blue tones (or vice versa)
indicates a ratchet reversal.
}
\label{fig:13}
\end{figure}

In Fig.~\ref{fig:13}(a,b), we plot heat maps of $\langle V_x\rangle$
and $\langle V_y\rangle$ as a function of $\rho$ versus $F_p$ for
the system from Fig.~\ref{fig:12}, highlighting both the forward and
reverse ratchet effects.
For $\rho < 0.83$, the ratchet motion
is primarily in the positive $x$ and negative $y$ directions,
with a reversal in the ratchet motion occurring at high densities.
It is possible that additional ratchet reversals could occur
for higher values of $\rho$ than those considered in this work.

\section{Thermal Effects}

\begin{figure}
\centering
\includegraphics[width=0.6\columnwidth]{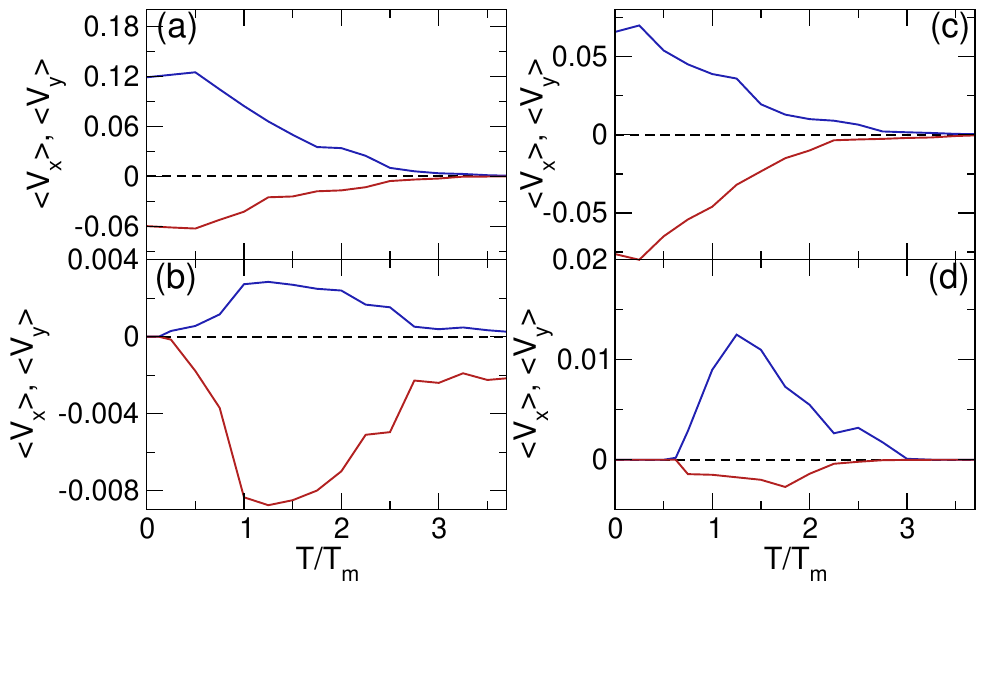}
\caption{
$\langle V_{x}\rangle$ (blue) and $\langle V_{y}\rangle$ (red)
vs temperature $T/T_m$  
in samples with $A_{p} = 0.75$, $\rho = 0.208$,
and $F_{AC} = 1.0$. Here $T_{m}$ is the temperature at which
the substrate-free system thermally melts.
(a,b) $x$ direction ac driving parallel to the substrate asymmetry direction
at (a) $qB/\alpha_d = 0.5$ and (b) $qB/\alpha_d = 3.0$.
(c,d) $y$ direction ac driving perpendicular to the
substrate asymmetry direction
at (c) $qB/\alpha_d= 1.0$ and
(d) $qB/\alpha_d = 0.4$.
}
\label{fig:14}
\end{figure}

We next consider the effect of adding thermal Langevin kicks to the
particle motion in order to represent a finite temperature.
We use a system with
$A_{p} = 0.75$, $\rho = 0.208$, and $F_{AC} = 1.0$, and we
report temperature in terms of
$T/T_{m}$, where $T_{m}$ is the temperature at which
the substrate-free system thermally disorders.
In Fig.~\ref{fig:14}(a) we plot
$\langle V_{x}\rangle$ and $\langle V_{y}\rangle$
versus $T/T_{m}$
at $qB/\alpha_d = 0.5$
for driving parallel to the substrate asymmetry direction.
There is initially a slight
increase in the ratchet effect with increasing $T/T_m$ followed by
a pronounced drop in ratchet motion for
$T/T_m > 1.0$.
The decrease occurs since the thermal fluctuations
reduce the effectiveness of the substrate,
and for $T/T_m > 1.0$, thermal hopping is strongly enhanced.
Figure~\ref{fig:14}(b) shows $\langle V_x\rangle$ and $\langle V_y\rangle$
versus $T/T_m$
for the same system at $qB/\alpha_d = 3.0$,
where for $T/T_m = 0$, the magnetic field is large enough to
localize the motion and prevent ratchet motion from occurring.
In this case, we find that the ratcheting motion reaches
a maximum for $T/T_m > 1.0$ and then diminishes;
however, the maximum velocity of the ratchet motion
is considerably reduced compared to that found at the lower value
of $qB/\alpha_d$.
In Fig.~\ref{fig:14}(c)
we illustrate $\langle V_{x}\rangle$ and
$\langle V_y\rangle$ versus $T/T_m$ for the same system
at $qB/\alpha_d=1.0$ but for driving perpendicular to the substrate
asymmetry direction. Here
the ratchet effect generally decreases with increasing $T/T_m$.
At $qB/\alpha_d=0.4$ in the same system, shown
in Fig.~\ref{fig:14}(d), no ratchet motion
occurs when $T/T_m = 0.0$;
however, for $T/T_m > 1.0$, thermal hopping of the particles out of the
substrate troughs produces a ratchet effect that is gradually destroyed
at high temperatures.
For $qB/\alpha_d > 3.0$, driving perpendicular
to the substrate asymmetry and driving parallel to the substrate
both produce similar behavior in which
there is finite ratchet motion that appears for $T/T_m > 1.0$
and then
diminishes with increasing temperature.
These results show that even for systems with strong thermal effects
or in a Wigner liquid, the transverse ratchet effects should be robust.

\section{Summary} 

We have examined ratchet effects for a two-dimensional system
of charged particles in a magnetic field interacting with a
one dimensional asymmetric substrate under ac driving applied
either parallel or perpendicular to the substrate asymmetry direction.
When the magnetic field is zero, a ratchet effect only occurs
when the ac drive is parallel to the substrate asymmetry direction.
Under a finite magnetic field, however,
a transverse ratchet effect can occur in which the particles move
both parallel and perpendicular to the substrate asymmetry direction
due to the finite Hall angle produced by the cyclotron motion
of the particles.
When the ac drive is applied perpendicular to the substrate asymmetry
direction, a transverse ratchet effect can occur.
For fixed ac driving amplitude, the ratchet effect
drops to zero at higher fields when
the charge motion becomes localized inside the pinning troughs.
We have investigated this ratchet effect
for varied magnetic fields, substrate strengths,
ac drive amplitudes, and charge density.
We find that there can be reversals of the ratchet effect
at higher charge densities, where the particles move against the
easy flow direction of the substrate asymmetry. The reversed ratchet
motion 
can occur for driving either parallel or perpendicular to the
substrate asymmetry direction.
Our results demonstrate that cyclotron motion and
the resulting finite Hall angle can produce new kinds of ratchet effects
that could be relevant for Wigner crystals, charged colloids,
dusty plasmas, and other charged systems coupled to
an asymmetric substrate
in the presence of a magnetic field.

\section*{Acknowledgements}
We gratefully acknowledge the support of the U.S. Department of
Energy through the LANL/LDRD program for this work.
This work was supported by the US Department of Energy through
the Los Alamos National Laboratory.  Los Alamos National Laboratory is
operated by Triad National Security, LLC, for the National Nuclear Security
Administration of the U. S. Department of Energy (Contract No. 892333218NCA000001).

\section*{References}

\bibliographystyle{iopart-num}
\bibliography{mybib}

\end{document}